\documentclass[pra,aps,twocolumn,superscriptaddress,showpacs]{revtex4}
\usepackage{graphicx}
\begin{document}

\title{Quantum correlation and entanglement between an ionizing system and a
neighbor atom interacting directly and via a quantized field}

\author{V. Pe\v{r}inov\'{a}}
\email{vlasta.perinova@upol.cz}
\author{A. Luk\v{s}}
\author{J. K\v{r}epelka}
\affiliation{Joint Laboratory of Optics, Palack\'{y} University, RCPTM,\\
17. listopadu 12, 77146 Olomouc, Czech Republic}
\author{J. Pe\v{r}ina, Jr.}
\affiliation{Institute of Physics of Academy of Sciences of the Czech
Republic,\\
Joint Laboratory of Optics, Palack\'{y} University, RCPTM,
17. listopadu 12, 77146 Olomouc, Czech Republic}

\date{}
\begin{abstract}
Quantum correlations between two neighbor atoms are studied. It is
assumed that one atomic system comprises a single auto-ionizing
level and the other atom does not contain any auto-ionizing level.
The excitation of both atoms is achieved by the interaction with
the same mode of the quantized field. It is shown that the
long-time behavior of two atoms exhibits quantum correlations even
when the atoms do not interact directly. This can be shown using
the optical excitation of the neighbor atom. Also a measure of
entanglement of two atoms can be applied after reduction of the
continuum to two levels.
\end{abstract}

\maketitle

\section{Introduction}

In the study of atoms with at least two electrons, bound states
and resonances are of interest. The resonances evolve into states
with one free electron after a very short time. This phenomenon is
called auto-ionization of the atom. With a revival of interest in
the auto-ionization, Fano published an appealing theoretical paper
\cite{F61} comprising an analysis of the excitation of the $2s2p$
level of helium by electrons. He argued that the natural line
shape contains a zero. Later, the optical absorption spectra of
the rare gases have been analyzed \cite{FC65}, while the paper
\cite{RB81} is one of many studies dealing with the mechanism of
atomic auto-ionization. A unified approach to the configuration
interaction and the influence of strong lasers have been expounded
in \cite{LaZ81}. In this framework, the studies \cite{RzE81,
LeTK87} have been realized. The quantum laser field has been taken
into account in \cite{LeBu90} and the effect of the squeezed state
has been studied in \cite{Le93}.

The Fano resonances can occur also in other physical settings. The
Fano resonances in nanoscale structures can be mentioned
\cite{MFlKi10}. The treatment of auto-ionization and the influence
of laser may be extended to a simultaneous auto-ionization, the
influence of laser, and to the interaction with a neighbor
two-level atom \cite{LPePKrLe10, PLLePe11a, PLLePe11b, PLPeLe11a,
PLPeLe11b}. The presence of a neighbor system may also
considerably increase photo-ionization and recombination rates
\cite{Najjari2010,Voitkiv2010}. In the analysis, the assumption of
weak optical pumping is usually used and leads to a simpler
behavior, cf., \cite{LaZ81}.

In \cite{LPLePe12}, the entanglement between an auto-ionization
system and a neighbor atom is studied for a classical driving
field. Besides the possibility to calculate a measure of
entanglement for the two atomic systems, a somewhat arbitrary, but
systematic, filtering is adopted. Two frequencies can be selected
in the auto-ionization system and the study of entanglement
reduces to the well-known two-qubit problem. In this paper, we
modify this analysis by including the quantal nature of the field.
In Sec.~II, we describe the model. In Sec.~III, we discuss
photoelectron spectra and the density plots of entanglement
measure. Sec.~IV provides conclusion.

\section{Quantum model of the optical excitation of two atoms}

We consider two mutually interacting atoms, $ a $ and $ b $, in
the presence of an electromagnetic field (for the scheme, see
Fig.~1). To quantize the electromagnetic field, we have to add to
the usual model annihilation and creation operators of the modes
which participate in the radiative interactions. Indeed, although
only the frequency $\omega_L$ of optical field is considered, an
infinity of modes at this frequency can be introduced.
\begin{figure}[tbp]
\center{
\includegraphics[width=7.0cm]{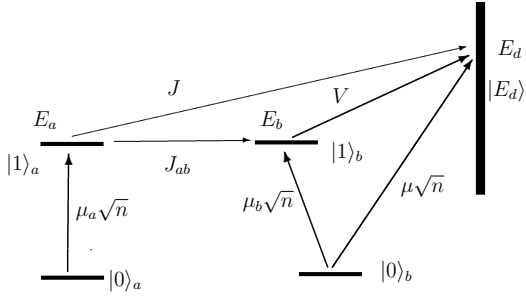}\hskip0mm
\caption{Sketch of an auto-ionization system $ b $ interacting
with a two-level atom $ a $ through a quantized field in the Fock
state $ |n\rangle $. The ground (excited) states are denoted as $
|0\rangle_a $ and $ |0\rangle_b $ ($ |1\rangle_a $, $ |1\rangle_b
$, and $ |E_{d}\rangle $).The dipole moments $ \mu_a $, $ \mu_b $,
and $\mu $ describe the interactions between atoms and field. The
excited discrete state of atom $ a $ ($ b $) has the energy $ E_a
$ ($ E_b $), whereas the energies $ E_{d} $ characterize the
excited states $ |E_{d}\rangle $ of the continuum. Symbol $ V $
stands for the Coulomb configurational coupling between the
excited states of atom $ b $. The constants $ J_{ab} $ and $ J $
describe the dipole-dipole interaction between the atoms $ a $ and
$ b $.} }
\end{figure}

We may suppose that the atom $a$ interacts with~the mode $L_a$ and
the atom $b$ interacts with the~mode $L_b$. We complete the levels
of the atomic system by the~photon-number states,
\begin{eqnarray}
&& \mbox{} |n_a,n_b\rangle_L\otimes|0,0\rangle_{ab}\mbox{,
}|n_a-1,n_b\rangle_L\otimes|1,0\rangle_{ab},
\nonumber\\
&& \mbox{} |n_a,n_b-1\rangle_L\otimes|0,1\rangle_{ab}\mbox{,
}|n_a-1,n_b-1\rangle_L\otimes|1,1\rangle_{ab},
\nonumber\\
&& \mbox{} |n_a,n_b-1\rangle_L\otimes|0,E_d\rangle_{ad},
\nonumber\\
&& \mbox{} |n_a-1,n_b-1\rangle_L\otimes|1,E_d\rangle_{ad},
\label{1}
\end{eqnarray}
where $n_a$ is a photon number in the mode $L_a$ and $n_b$ is a
photon number in the mode $L_b$. In Eq.~(\ref{1}), $|0\rangle_a$
($|0\rangle_b$) is the ground state of the atom $a$ ($b$),
$|1\rangle_a$ is the excited state of the atom $a$, $|1\rangle_b$
is the auto-ionization state of the atom $b$, $|E_d\rangle$
$\equiv$ $|E_d\rangle_d$ is the continuum state of the atom $b$,
and $E_d$ is an energy difference between the ground state
$|0\rangle_b$ and the state $|E_d\rangle$. Here we have used the
photon-number states $|n_a\rangle$, $|n_b\rangle$,
$|n_a-1\rangle$, and $|n_b-1\rangle$ simultaneously to indicate
that the Hilbert space of the states can be decomposed into
invariant subspaces. For $n_a,n_b\ge1$, these subspaces have a
dimension equal to 6. Each invariant subspace is a tensorial
product of the subspaces corresponding to the Jaynes--Cummings
model (dimension 2) and the model due to Leo\'{n}ski and Bu\v{z}ek
(dimension 3) \cite{LeBu90}. The Hamiltonian has the form
\begin{equation}
\hat{H}=\hat{H}^{'}_{\rm free}
+\hat{H}^{'}_{\rm a-i}+\hat{H}^{}_{\rm t-a}+\hat{H}^{}_{\rm trans},
\label{2}
\end{equation}
where
\begin{equation}
\hat{H}^{'}_{\rm free}=\hbar\omega_L(\hat{a}_L^\dagger\hat{a}_L +
\hat{b}_L^\dagger\hat{b}_L),
\label{3}
\end{equation}
with $\hat{a}_{L}$ and $\hat{b}_{L}$  ($\hat{a}_{L}^{\dagger}$ and
$\hat{b}_{L}^{\dagger}$) being the photon annihilation (creation)
operators. The Hamiltonian $ \hat{H}^{'}_{\rm a-i} $ of atom $ b $
with auto-ionizing level in Eq.~(\ref{2}) is written as
\begin{eqnarray}
&& \mbox{} \hat{H}^{'}_{\rm a-i}=E_b|1\rangle_b{}_b\langle1|+
\int E_d|E_d\rangle\langle E_d|\,dE_d
\nonumber\\
&& \mbox{} + \int \left(V|E_d\rangle\,{}_b\langle1|+\mbox{H.c.}\right)\,dE_d
\nonumber\\
&& \mbox{} + \left(\mu_{b}\hat{b}_{L}|1\rangle_b{}_b\langle0|+\mbox{H.c.}\right)
\nonumber\\
&& \mbox{} + \int \left(\mu\hat{b}_{L}|E_d\rangle\,{}_b\langle0|+
\mbox{H.c.}\right)\,dE_d,
\label{4}
\end{eqnarray}
where $E_b$ means an energy difference between the ground state
$|0\rangle_b$ and the state $|1\rangle_b$. Symbol $\mu_{b}$ gives
the strength of optical excitation from the ground state
$|0\rangle_{b}$ into the auto-ionization state $|1\rangle_{b}$,
$\mu$ is the strength of optical excitation from the ground state
$|0\rangle_{b}$ of the atom $b$ into the continuum state
$|E_{d}\rangle$, and $V$ describes the Coulomb configuration
interaction between the excited states of atom $b$. The
Hamiltonian of the neighbor two-level atom in Eq.~(\ref{2}) reads
\begin{equation}
\hat{H}_{\rm t-a}=E_a|1\rangle_a{}_a\langle1|+
\left(\mu_{a}\hat{a}_{L}|1\rangle_a{}_a\langle0|+\mbox{H.c.}\right),
\label{5}
\end{equation}
where $E_a$ means an energy difference between the ground state $|0\rangle_a$
and the state $|1\rangle_a$, $\mu_{a}$ is the strength of optical excitation
from the ground state $|0\rangle_{a}$ into the excited state $|1\rangle_{a}$.

In Eq.~(\ref{2}), the Hamiltonian $\hat{H}_{\rm trans}$
characterizes the dipole--dipole interaction between the atoms $a$
and $b$,
\begin{eqnarray}
&& \mbox{}  \hat{H}_{\rm trans}=
\left(J_{ab}|1\rangle_b{}_b\langle0||0\rangle_a{}_a\langle1|+\mbox{H.c.}
\right)
\nonumber\\
&& \mbox{} + \int \left(J|E_d\rangle\,{}_b\langle0||0\rangle_a{}_a\langle1|+
\mbox{H.c.}\right)\,dE_d,
\label{6}
\end{eqnarray}
where $J_{ab}$ $(J)$ characterize energy transfer from the ground
state $|0\rangle_b$ into the state $|1\rangle_b$ $(|E_{d}\rangle)$
at the cost of the decay from the state $|1\rangle_a$ into the
state $|0\rangle_a$. We note that if $J_{ab}=0$ and $J=0$, the
Hamiltonian $\hat{H}$ describes uncoupled atoms.

We will treat the situation where the atoms $a$ and $b$
interact with~a single mode $L$, $\hat{b}_{L}\rightarrow
\hat{a}_{L}$. In this case, the levels written in Eq.~(\ref{1})
simplify,
\begin{eqnarray}
&& \mbox{} |n\rangle_L\otimes|0,0\rangle_{ab}\mbox{,
}|n-1\rangle_L\otimes|1,0\rangle_{ab},
\nonumber\\
&& \mbox{} |n-1\rangle_L\otimes|0,1\rangle_{ab}\mbox{,
}|n-2\rangle_L\otimes|1,1\rangle_{ab},
\nonumber\\
&& \mbox{} |n-1\rangle_L\otimes|0,E_d\rangle_{ad}\mbox{,
}|n-2\rangle_L\otimes|1,E_d\rangle_{ad},
\label{7}
\end{eqnarray}
and $n$ is the number of photons in the mode $L$. Here we have
used the photon-number states $|n\rangle_L$, $|n-1\rangle_L$, and
$|n-2\rangle_L$ simultaneously to indicate that the Hilbert space
of the states can be decomposed into invariant subspaces. But in
the~case of a single mode, an invariant subspace cannot be
investigated as a tensorial product. We can see from
Eq.~(\ref{7}), that the atom $a$ at the level $|0\rangle_a$
interacts with the~field in the state $|n\rangle_L$ or
$|n-1\rangle_L$ in the dependence on the state of the atom $b$ and
the atom $a$ at the level $|1\rangle_a$ interacts with the~field
in the state $|n-1\rangle_L$ or $|n-2\rangle_L$ in the~dependence
on the state of the atom $b$.

The Hamiltonian has the form
\begin{equation}
\hat{H}=\hat{H}_{\rm free}+\hat{H}_{\rm a-i}+\hat{H}_{\rm t-a}+
\hat{H}_{\rm trans}
\label{8}
\end{equation}
where
\begin{equation}
\hat{H}_{\rm free}=\hbar\omega_L\hat{a}_L^\dagger\hat{a}_L
\label{9}
\end{equation}
and
\begin{equation}
\hat{H}_{\rm a-i}=\left.\hat{H}'_{\rm a-i}\right|_{\hat{b}_L
\rightarrow\hat{a}_L}.
\label{10}
\end{equation}
Following \cite{LeBu90}, we modify the Schr\"{o}dinger picture by
considering the state vector $|\psi\rangle(t)$ in the form
\begin{eqnarray}
&& \mbox{} |\psi\rangle(t)={\sum_{n=0}^{\infty}}'
\exp\left[-\frac{i}{\hbar}E_L(n-2)t\right]
\nonumber\\
&& \mbox{} \times \Bigl[c_{00}^{(n)}(t)|n\rangle_L|0,0\rangle_{ab}+
c_{10}^{(n-1)}(t)|n-1\rangle_L|1,0\rangle_{ab}
\nonumber\\
&& \mbox{} + c_{01}^{(n-1)}(t)|n-1\rangle_L|0,1\rangle_{ab}
\nonumber\\
&& \mbox{} + c_{11}^{(n-2)}(t)|n-2\rangle_L|1,1\rangle_{ab}
\nonumber\\
&& \mbox{} + \int d_0^{(n-1)}(E_d,t)|n-1\rangle_L|0,E_{d}\rangle_{ad}\,dE_d
\nonumber\\
&& \mbox{} + \int d_1^{(n-2)}(E_d,t)|n-2\rangle_L|1,E_d\rangle_{ad}\,dE_d\Bigr],
\label{11}
\end{eqnarray}
where $E_{L}=\hbar \omega_{L}$. The prime indicates that for
$n=0,1$ some of the components must be omitted. The components
with $|n-2\rangle_L $ have to be omitted for $n=0,1$ and those
with $|n-1\rangle_L $ have to be left out for $n=0$.

It holds that $[\hat{J},\hat{H}]$ $=$ $\hat{0}$ for
\begin{equation}
\hat{J}=|1\rangle_b\,{}_b\langle1|+
\int|E_d\rangle\langle E_d|\,dE_d
+ |1\rangle_a\,{}_a\langle1|+
\hat{a}^\dagger_{L}\hat{a}_{L}.
\label{12}
\end{equation}
An invariant subspace ${\cal H}^{(n)}$ is the eigenspace of the operator
$\hat{J}$ related to an eigenvalue $n$. We assume that $n\geq 2$. In this
invariant subspace, the composite system is described by the equations
\begin{eqnarray}
 \frac{d}{dt}{\bf c}^{(n)}(t) &=& - \frac{i}{\hbar}{\bf A}^{(n)} {\bf
  c}^{(n)}(t) \nonumber\\
  & & \mbox{} - \frac{i}{\hbar}\int{\bf B}_1^{(n)}{\bf d}^{(n)}(E_d,t)\,dE_d,
   \nonumber\\
 \frac{d}{dt}{\bf d}^{(n)}(E_d,t) &=& - \frac{i}{\hbar}{\bf B}_2^{(n)}
  {\bf c}^{(n)}(t) \nonumber\\
 & & \mbox{} - \frac{i}{\hbar}{\bf K}^{(n)}(E_d){\bf d}^{(n)}(E_d,t),
\label{13}
\end{eqnarray}
where
\begin{eqnarray}
&& \mbox{} {\bf c}^{(n)}(t)=\left(\begin{array}{c}
c_{00}^{(n)}(t)\\c_{10}^{(n-1)}(t)\\c_{01}^{(n-1)}(t)\\c_{11}^{(n-2)}(t)
\end{array}\right)\mbox{, }
\nonumber\\
&& \mbox{} {\bf d}^{(n)}(E_d,t)=\left(\begin{array}{c}
d_{0}^{(n-1)}(E_d,t)\\d_{1}^{(n-2)}(E_d,t)
\end{array}\right).
\label{14}
\end{eqnarray}

Further
\begin{equation}
{\bf A}^{(n)}=\left(\begin{array}{cccc}
2E_L&\mu_{a}^*\sqrt{n}&\mu_{b}^*\sqrt{n}&0\\
\mu_{a}\sqrt{n}&E_a+E_L&J_{ab}^*&\mu_{b}^*\sqrt{n-1}\\
\mu_{b}\sqrt{n}&J_{ab}&E_b+E_L&\mu_{a}^*\sqrt{n-1}\\
0&\mu_{b}\sqrt{n-1}&\mu_{a}\sqrt{n-1}&E_a+E_b
\end{array}\right),
\label{15}
\end{equation}
\begin{eqnarray}
&& \mbox{} {\bf B}_1^{(n)}=\left(\begin{array}{cc}
\mu_{}^*\sqrt{n}&0\\
J_{}^*&\mu_{}^*\sqrt{n-1}\\
V_{}^*&0\\
0&V_{}^*
\end{array}\right),
\label{16}\\
&& \mbox{} {\bf B}_2^{(n)}={\bf B}_1^{(n)\dagger},
\label{17}\\
&& \mbox{} {\bf K}^{(n)}(E_d)=\left(\begin{array}{cc}
E_d+E_L&\mu_{a}^*\sqrt{n-1}\\
\mu_{a}\sqrt{n-1}&E_a+E_d
\end{array}\right).
\label{18}
\end{eqnarray}
We introduce the matrix
\begin{equation}
{\bf M}^{(n)}={\bf A}^{(n)}-i\pi{\bf B}_1^{(n)}{\bf B}_2^{(n)},
\label{19}
\end{equation}
and let $\xi_1^{(n)}$ and $\xi_2^{(n)}$ denote the eigenvalues of the matrix
${\bf K}^{(n)}(0)$ and
$\Lambda^{(n)}_{{\bf M}^{(n)}j}$, $j=1,2,3,4$, be the eigenvalues of the
matrix ${\bf M}^{(n)}$. Let us recall the possibility of decompositions
\begin{eqnarray}
{\bf K}^{(n)}(0) &=&\xi_1^{(n)}{\bf K}^{(n)}_1+\xi_2^{(n)}{\bf K}^{(n)}_2,
\label{20}\\
{\bf M}^{(n)} &=& \sum_{j=1}^4\Lambda^{(n)}_{{\bf M}^{(n)}j}{\bf M}^{(n)}_{j},
\label{21}
\end{eqnarray}
where ${\bf K}_1^{(n)}$, ${\bf K}_2^{(n)}$ are solutions of the equations
\begin{eqnarray}
{\bf K}_1^{(n)}+{\bf K}_2^{(n)} &=& {\bf I}_2,
\nonumber\\
\xi_1^{(n)}{\bf K}_1^{(n)}+\xi_2^{(n)}{\bf K}_2^{(n)} &=& {\bf
K}^{(n)}(0). \label{22}
\end{eqnarray}
Similarly,  ${\bf M}_j^{(n)}$, $j=1,2,3,4$, are solutions of the
equations
\begin{equation}
\sum_{j=1}^{4} \Lambda^{(n)k}_{{\bf M}^{(n)}j}{\bf M}^{(n)}_{j}
={\bf M}^{(n)k},\;\;k=0, 1, 2, 3.
\label{23}
\end{equation}
In Eqs.~(\ref{22}) and (\ref{23}), ${\bf I}_2$ and $\sum_{j=1}^{4}
{\bf M}^{(n)}_{j} $ $=$ ${\bf I}_4$ are $2\times2$ and $4\times4$
unit matrices, respectively.

The first vector of the components of the solution of
Eqs.~(\ref{13}) has the very simple form
\begin{equation}
{\bf c}^{(n)}(t)= \exp\left( -\frac{i}{\hbar}{\bf M}^{(n)}t\right)
{\bf c}^{(n)}(0). \label{24}
\end{equation}

We introduce a $2\times4$ matrix ${\bf T}^{(n)}(E_d)$ as the solution of the
Sylvester equation
\begin{equation}
{\bf K}^{(n)}(E_d){\bf T}^{(n)}(E_d)-{\bf T}^{(n)}(E_d){\bf M}^{(n)}=
{\bf B}_2^{(n)}.
\label{25}
\end{equation}

The solution has the form
\begin{eqnarray}
&& {\bf T}^{(n)}(E_d)=\sum_{k=1}^2\sum_{j=1}^4 \frac{ {\bf
K}_k^{(n)} {\bf B}^{(n)}_2{\bf M}_j^{(n)}
}{E_d+\xi_k^{(n)}-\Lambda_{{\bf M}^{(n)}j}} . \label{26}
\end{eqnarray}
The dependence of the components of the amplitude spectrum on the
initial state of the system with~${\bf d}^{(n)}(E_d,0)={\bf0}$ is
\begin{eqnarray}
&& \mbox{} {\bf d}^{(n)}(E_d,t)=\Biggl\{ \exp \left[
-\frac{i}{\hbar}{\bf K}^{(n)}(E_d)t\right] {\bf T}^{(n)}(E_d)
\nonumber\\
&& \mbox{} - {\bf T}^{(n)}(E_d) \exp\left[-\frac{i}{\hbar}{\bf
M}^{(n)}t \right] \Biggr\}{\bf c}^{(n)}(0). \label{27}
\end{eqnarray}
We observe that
\begin{equation}
{\bf d}^{(n)}(E_d,t)\simeq{\bf d}_{\rm
out}^{(n)}(E_d,t)\;\;\mbox{for}\;\;t \rightarrow\infty,
\label{28}
\end{equation}
where
\begin{equation}
{\bf d}_{\rm out}^{(n)}(E_d,t)= \exp\left[-\frac{i}{\hbar}{\bf
K}^{(n)}(E_d)t \right] {\bf T}^{(n)}(E_d){\bf c}^{(n)}(0).
\label{29}
\end{equation}

The increase of the diagonal terms by~$2E_L$ means that also the
eigenvalues $\xi_k^{(n)}$, $k=1,2$, $\Lambda_{{\bf M}j}^{(n)}$ are
raised by the~same amount. In the relations like (\ref{26}), these
increments mutually cancel and elsewhere they already correspond
to the relation (\ref{11}).

All the previous exposition beginning with (\ref{14}) should be
modified for $n<2$. Let us note that the initial vacuum
field and the ground states of the atoms $a$ and $b$,
$n=0$, do not lead to any transitions to the continuum
states. It holds that ${\bf c}^{(0)}(t)$ $=$
$(c_{00}^{(0)}(t))$ and the description reduces to the
equation $\frac{d}{dt}c_{00}^{(0)}(t)$ $=$
$-\frac{i}{\hbar}2E_Lc_{00}^{(0)}(t)$. The reduction for
$n=1$ is a consequence of non-existence of $|n-2\rangle_L$
as in (\ref{11}) and need not be made explicit. Let us note that
transition to a continuum state can occur for $n=1$, but not
simultaneously with an excitation of the atom $a$.

\section{Numerical results}

The long-time behavior is characterized by a complete ionization
of the atom with an auto-ionizing level and by both the levels of
the two-level atom $ a $ being occupied. The long-time behavior is
periodical due to the dynamics of the two-level atom in the cw
laser field. At all times the spectra can be determined as the
probability distribution of the two-level atom at its levels and
of the atom with the auto-ionizing level in the continuum of
levels. By the normalization, conditional spectra are defined. The
difference between the conditional spectra is an effect of the
atomic quantum correlation. The difference between the conditional
spectra can be seen even in the case where the dipole--dipole
interaction of the atoms is missing.

An important case of the quantum correlation is the entanglement.
We measure this entanglement using the negativity. The
entanglement is conserved, even though we restrict the continuum
of levels to two of them, in the most of the pairs of the selected
frequencies. We calculate the negativity of the partially
transposed statistical matrix for two levels of the two-level atom
and selected continuum levels of the atom with an auto-ionizing
level.

\subsection{Appropriate parametrization}

In the previous work, the functions
\begin{equation}
q_b=\frac{\mu_b}{\pi V^*\mu}\mbox{, }\gamma_b=\pi|V|^2
\label{30}
\end{equation}
of the parameters of the atom $b$ without a neighbor have been
conveniently introduced. Also the functions
\begin{equation}
q_a=\frac{\mu_a}{\pi J^*\mu}\mbox{, }\gamma_a=\pi|J|^2
\label{31}
\end{equation}
of the parameters of both the atoms have been defined. Here we
conveniently introduce a one-photon version of the usual
'excitation' parameter $ \Omega $,
\begin{equation}
\Omega=\sqrt{4\pi\Gamma(Q^2+1)}\mu,
\label{32}
\end{equation}
where
\begin{equation}
\Gamma=\gamma_a+\gamma_b\mbox{, }
Q=\frac{\gamma_aq_a+\gamma_bq_b}{\Gamma}.
\label{33}
\end{equation}
By the replacements $\mu_b\rightarrow J_{ab}$,
$\mu\rightarrow J$ in the function $q_b$, the function
\begin{equation}
q_{\rm trans}=\frac{J_{ab}}{\pi V^*J}
\label{34}
\end{equation}
originates. For $V,J\ge0$, the parameters of the model can be
expressed in the forms
\begin{eqnarray}
&& \mbox{} V=\sqrt{\frac{\gamma_b}{\pi}}\mbox{, }
J=\sqrt{\frac{\gamma_a}{\pi}}\mbox{, }
\mu=\frac{\Omega}{\sqrt{4\pi\Gamma(Q^2+1)}},
\nonumber\\
&& \mbox{} \mu_a=\pi J^*\mu q_a\mbox{, } \mu_b=\pi V^*\mu
q_b\mbox{, } J_{ab}=\pi V^*Jq_{\rm trans}. \hspace{5mm}
\label{35}
\end{eqnarray}
From this, $q_a,\gamma_a,q_b,\gamma_b,\Omega,q_{\rm trans}$ are
new parameters.

In what follows, we will assume $E_a=E_b=E_L=1$ and four different
physically interesting cases that elucidate the behavior of the
analyzed system:

\noindent (a) $q_a=0$; $\gamma_a=0$; $q_b=\gamma_b=1$;
$\Omega=0.1,1$; $q_{\rm trans}=0$,

\noindent (a') $q_a=100$; $\gamma_a=0$; $q_b=\gamma_b=1$;
$\Omega=0.1,1$; $q_{\rm trans}=0$,

\noindent (b) $q_a=100$; $\gamma_a=10^{-4}$; $q_b=\gamma_b=1$;
$\Omega=0.1$; $q_{\rm trans}=0$,

\noindent (c) $q_a=100$; $\gamma_a=10^{-4}$; $q_b=\gamma_b=1$;
$\Omega=0.1$; $q_{\rm trans}=1$.

Whereas atom $ b $ is alone in (a), it feels the presence of atom
$ a $ due to the quantized optical field in (a'). In (b), both
atoms interact by the dipole-dipole interaction that includes only
the continuum of states at the atom $ b $. Finally, also the
dipole-dipole interaction between the discrete levels of both
atoms is taken into play in (c). We note that detuning of energy
levels of both atoms from the laser frequency does not
qualitatively modify the behavior of the system (for more details,
see \cite{LPLePe12} for semiclassical model). Also, we analyze the
system at time $ t=2 $ bellow. For the considered values of
parameters, the behavior of the system at time $ t=2 $ already
corresponds to that appropriate to the long-time limit.

\subsection{The role of atom $ a $ in forming the ionization spectra}

In the model, atoms $ a $ and $ b $ are in fact mutually coupled
by two types of interactions. Side by side with the discussed
dipole-dipole interaction, the interaction mediated by photons in
the quantized field also occurs. This interaction qualitatively
distinguishes the presented fully quantum model from the common
semiclassical models that assume a classical predefined optical
pump field \cite{PLLePe11a, PLLePe11b}.

The long-time limit of the two atomic systems is described by a
statistical matrix $(\rho_{jk}^{\rm out}(E_d,E'_d,t))$, with two
discrete indices $j,k$ and two continuous arguments $E_d$, $E'_d$.
Here
\begin{eqnarray}
&& \mbox{} \rho_{jk}^{\rm out}(E_d,E'_d,t)
\nonumber\\
&& \mbox{} =\sum_{n=\max(1+j,1+k)}^{\infty} d_{j{\rm
out}}^{(n-1-j)}(E_d,t)d_{k{\rm out}}^{(n-1-k)*}(E'_d,t).
\nonumber\\
\label{36}
\end{eqnarray}
As usual, the photoelectron spectra are identified with the distributions
\begin{equation}
W_{j}^{\rm out}(E_d,t)=\rho_{jj}^{\rm out}(E_d,E_d,t)\mbox{,
}j=0,1.
\label{37}
\end{equation}
This joint description may be reduced to the marginal probability
distribution of the levels of the atom $a$,
\begin{equation}
p_{j}^{\rm out}(t)= \int_{-\infty}^\infty W_{j}^{\rm
out}(E_d,t)dE_d.
\label{38}
\end{equation}
We consider also the conditional distributions or spectra
\begin{equation}
W_{|j}^{\rm out}(E_d,t)= \frac{W_{j}^{\rm out}(E_d,t)}{p_j^{\rm
out}(t)}\mbox{, }j=0,1.
\label{39}
\end{equation}
The closed formula for $p_{j}^{\rm out}(t)$ is rather complicated,
\begin{equation}
p_{j}^{\rm out}(t)= \sum_{n=1+j}^\infty\int_{-\infty}^\infty
\left|d_{j\rm out}^{(n-1-j)}(E_d,t)\right|^2dE_d,
\label{40}
\end{equation}
where
\begin{eqnarray}
&& \mbox{} \int_{-\infty}^\infty\left|d_{j\rm out}^{(n-1-j)}(E_d,t)
\right|^2dE_d
\nonumber\\
&& \mbox{} = \left( \int_{-\infty}^\infty {\bf d}_{\rm
out}^{(n)}(E_d,t){\bf d}_{\rm out}^{(n)\dagger}(E_d,t)d E_d
\right)_{jj},
\label{41}
\end{eqnarray}
with
\begin{eqnarray}
&& \mbox{} \int_{-\infty}^\infty
{\bf d}_{\rm out}^{(n)}(E_d,t){\bf d}_{\rm out}^{(n)\dagger}(E_d,t)d E_d
= 2\pi
\nonumber\\
&& \mbox{}\times \sum_{k=1}^2\sum_{j=1}^4
\sum_{k'=1}^2\sum_{j'=1}^4
\frac{\exp\left[\frac{i}{\hbar}(\xi_{k'}^{(n)}-\xi_k^{(n)})t
\right]} {i\left(\xi_{k'}^{(n)}-\xi_k^{(n)}- \Lambda_{{\bf
M}^{(n)}j'}^*+\Lambda_{{\bf M}^{(n)}j}\right)}
\nonumber\\
&& \mbox{} \times {\bf K}_k^{(n)}{\bf B}_2^{(n)}{\bf M}_j^{(n)}
{\bf c}^{(n)}(0){\bf c}^{(n)\dagger}(0) {\bf
M}_{j'}^{(n)\dagger}{\bf B}_1^{(n)}{\bf K}_{k'}^{(n)}.
\label{42}
\end{eqnarray}
The long-time total photoelectron spectrum is time-independent,
\begin{equation}
W^{\rm out}(E_d)=W_0^{\rm out}(E_d,t)+W_1^{\rm out}(E_d,t).
\label{43}
\end{equation}
\begin{figure}[tbp]
\center{
\includegraphics[width=7.0cm]{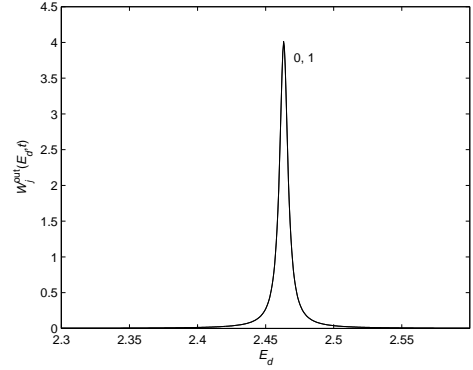} \hskip0mm
\caption{Unconditioned photoelectron spectra $W_{j}^{\rm
out}(E_d,t)$, $t=2$, $j=0,1$. Initially the laser mode is in a
coherent state with the mean photon number equal to 1. The photon
energy is $E_L=1$. The energy differences $E_b=E_a=1$. The
parameters $q_a=100$, $\gamma_a=0$, $q_b=\gamma_b=1$,
$\Omega=0.1$, $q_{\rm trans}=0$.} }
\end{figure}

\begin{figure}[tbp]
\center{
\includegraphics[width=7.0cm]{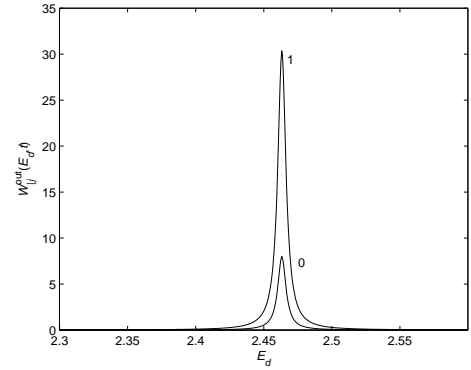}\hskip0mm
\caption{Same as in Fig.~2, but the conditional spectra
$W_{|j}^{\rm out}(E_d,t)$, $ j=0,1 $, are plotted.} }
\end{figure}

In Figs.~2 and~3 the case (a') of data with $\Omega=0.1$ and
for the initial coherent state $|1\rangle_L$ is illustrated.
The unconditioned and conditional photoelectron spectra have
a multi-peak structure and the peak positions are about the
same for both the values of the subscript $j$. Therefore the
plot is restricted to an interval which includes a single
peak of a spectrum.
In Fig.~2, it is seen that the unconditioned photoelectron
spectra coincide and cannot be discerned in the chosen
interval. In contrast, in Fig.~3 it is obvious that the
conditional spectra differ significantly in the selected
interval. It proves the dependence of the occupation of
the atom $b$'s level on the atom $a$'s level.

To reveal the role of atom $ a $ in ionization of atom $ b $, we
compare the long-time ionization spectra of atom $ b $ for atom $
a $ present and absent. We consider a greater value of
single-photon Rabi frequency $ \Omega $ to emphasize quantum
features of the model ($ \Omega = 1 $).

\begin{figure}[tbp]
\center{
\includegraphics[width=7.0cm]{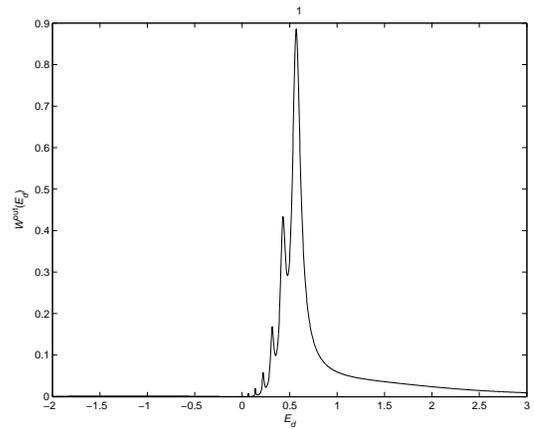}\hskip0mm
\caption{Photoelectron spectrum $W^{\rm out}(E_d)$. Initially the
laser mode is in a coherent state with the mean photon number
equal to 1. The parameters $E_a=E_b=E_L=1$, $q_a=0$, $\gamma_a=0$,
$q_b=\gamma_b=1$, $\Omega=1$, $q_{\rm trans}=0$.} }
\end{figure}

\begin{figure}[tbp]
\center{
\includegraphics[width=7.0cm]{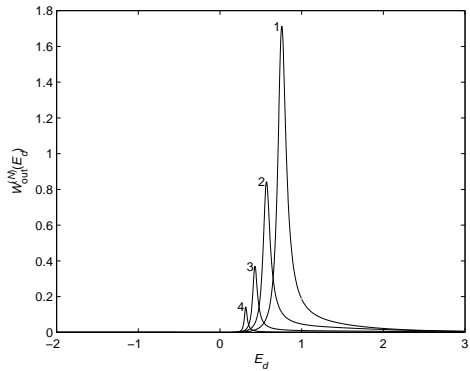}\hskip0mm
\caption{Fock components $W_{\rm out}^{(N)}(E_d)$, $N=1,...,4$,
of the spectrum $W^{\rm out}(E_d)$. The parameters are the same as in
Fig.~4.}
}
\end{figure}
Ionization of isolated atom $ b $ in a quantized field leads, in
general, to the occurrence of sharp peaks in the ionization
spectra (see Fig.~4). These peaks arise from the ionization caused
by individual Fock states of the optical field. This is documented
in Figs.~4, 5, in which the ionization spectra corresponding to
the coherent and Fock states are shown. It holds that the greater
the Fock number $ n $ is, the narrower is the corresponding
spectral peak and also the closer is the peak to the position of
energy of the Fano zero (see Fig.~4). Such behavior qualitatively
resembles that of an ionization spectrum caused by a classical
strong pump field \cite{RzE81}.

\begin{figure}[tbp]
\center{
\includegraphics[width=7.0cm]{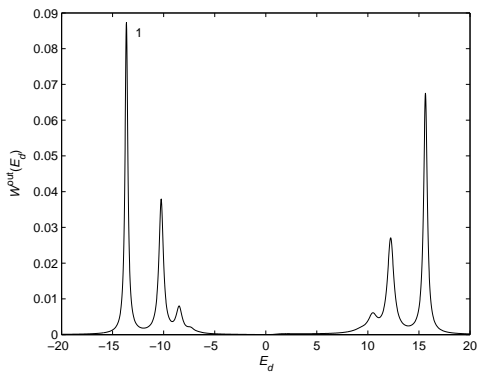}\hskip0mm
\caption{Photoelectron spectrum $W^{\rm out}(E_d)$. The parameters are the same
as in Fig.~4, but the parameter $q_a=100$.}
}
\end{figure}

\begin{figure}[tbp]
\center{
\includegraphics[width=7.0cm]{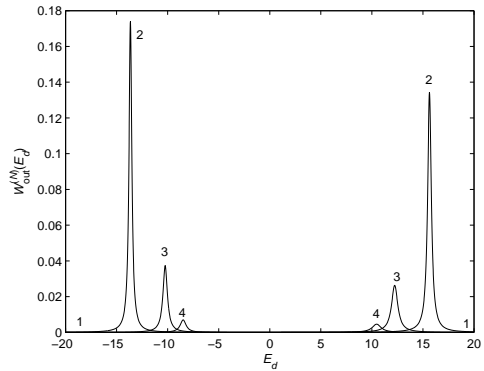}\hskip0mm
\caption{Fock components $W_{\rm out}^{(N)}(E_d)$, $N=1,...,4$,
of the spectrum $W^{\rm out}(E_d)$. The parameters are the same
 as in Fig.~4, but the parameter $q_a=100$.}
}
\end{figure}

The presence of atom $ a $ in the quantized pump field
considerably modifies the ionization spectra of atom $ b $ due to
the mutual indirect interaction of both atoms through the
quantized pump field. Contrary to the spectra of isolated atom $ b
$, the ionization spectra have contributions both below and above
the pump-field frequency. Moreover, the spectral peaks above and
below the pump-field frequency occur in pairs which results in
nearly symmetric ionization spectra (see Figs.~6,~7). This
symmetry is inherent to the Fock states from which it transfers
into the coherent states, as documented in Figs.~6,~7. It holds
that the greater the Fock number $ n $, the closer the two peaks
to the pump-field frequency.

\begin{figure}[tbp]
\center{
\includegraphics[width=6.0cm]{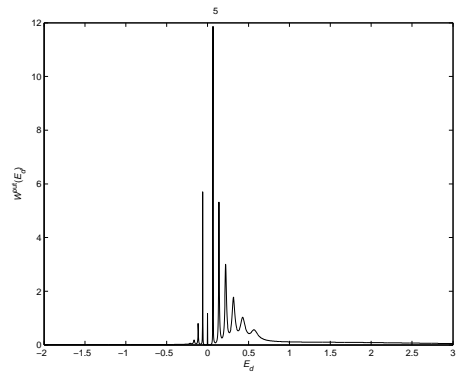}\hskip0mm
\caption{Photoelectron spectra $W^{\rm out}(E_d)$. The parameters
$E_a=E_b=E_L=1$, $\gamma_a=0$, $q_b=\gamma_b=1$, $\Omega=1$, $q_{\rm trans}=0$.
Here the initial mean photon number is equal to 5, $q_a=0$.}
}
\end{figure}

\begin{figure}[tbp]
\center{
\includegraphics[width=6.0cm]{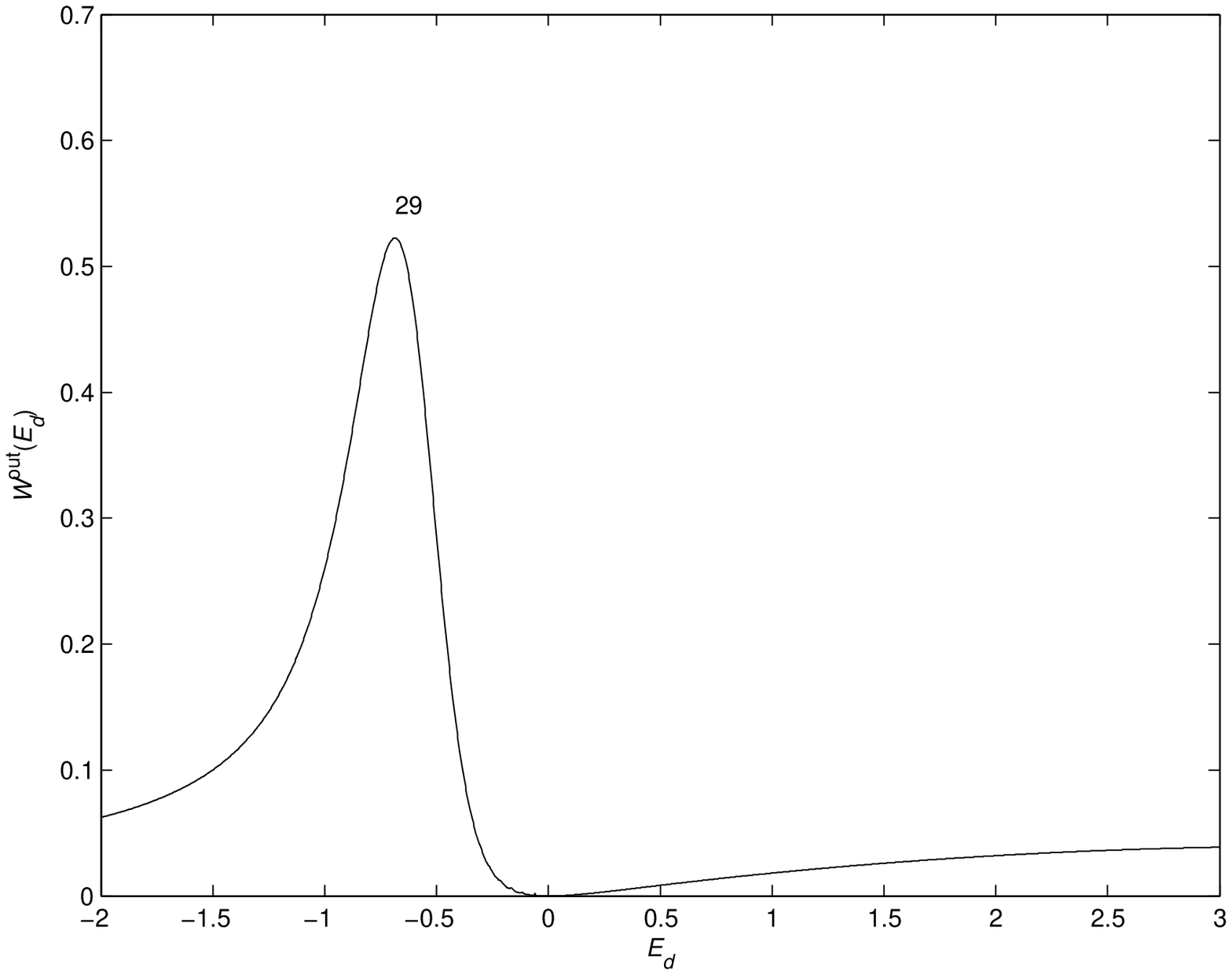}\hskip0mm
\caption{Same as in Fig.~8, but the initial mean photon number is equal to
29, $q_a=0$.}
}
\end{figure}

\begin{figure}[tbp]
\center{
\includegraphics[width=7.0cm]{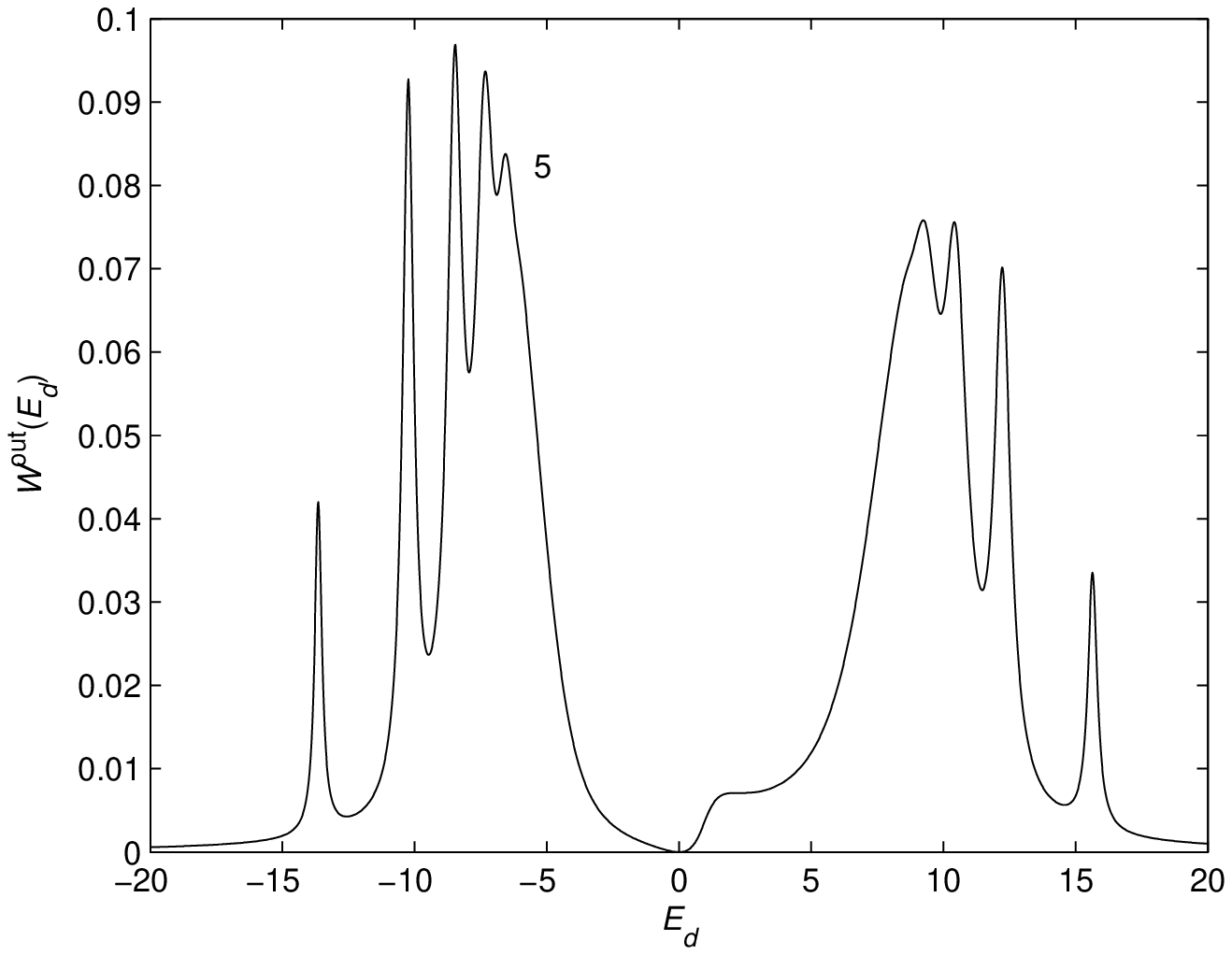}\hskip0mm
\caption{Same as in Fig.~8, but the initial mean photon number is equal to 5,
$q_a=100$.}
}
\end{figure}

\begin{figure}[tbp]
\center{
\includegraphics[width=7.0cm]{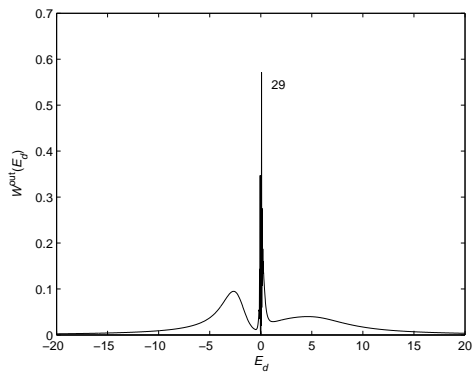}\hskip0mm
\caption{Same as in Fig.~8, but the initial mean photon number is equal to
29, $q_a=100$.}
}
\end{figure}

If the pump-field intensity increases, the spectrum of isolated
atom $ b $ is built more and more from contributions of
higher-number Fock states and it moves to lower energies crossing
the energy of Fano zero. The more intense the pump field, the more
suppressed (smoothed) the spectral structure of individual Fock
states (see Figs.~8,~9). Also the narrowing of the overall
ionization spectrum in the vicinity of the energy of Fano zero is
observed. When atom $ a $ is present, the ionization spectra also
gradually lose their peaked structure with the increasing
pump-field intensities (see Figs.~10,~11). For sufficiently high
pump-field intensities, the spectrum approaches that of the
isolated atom $ a $.

\begin{figure}[tbp]
\center{
\includegraphics[width=7.0cm]{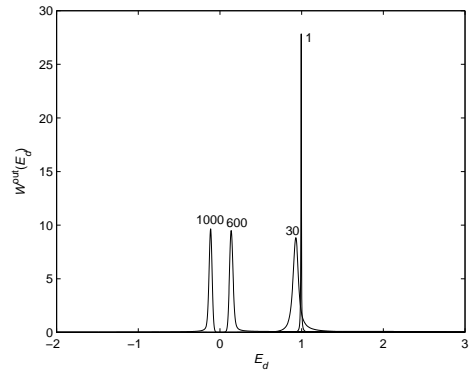}\hskip0mm
\caption{Photoelectron spectra $W^{\rm out}(E_d)$. Initially the laser
mode is in coherent states with the mean photon numbers
equal to 1, 30, 600, 1000. The parameters $E_a=E_b=E_L=1$,
$\gamma_a=0$, $q_b=\gamma_b=1$, $\Omega=0.1$, $q_{\rm trans}=0$. Here $q_a=0$.}
}
\end{figure}

\begin{figure}[tbp]
\center{
\includegraphics[width=7.0cm]{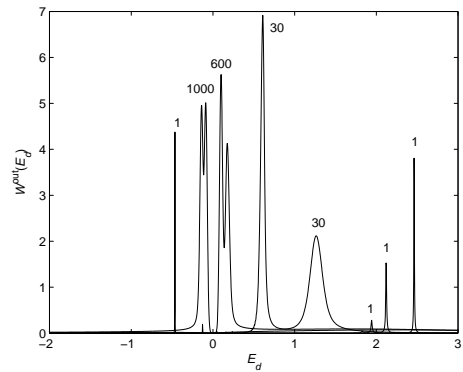}\hskip0mm
\caption{Same as in Fig.~12, but $q_a=100$.}
}
\end{figure}

When the interaction mediated by the quantized field is weaker,
the behavior of ionization spectra with the increasing pump-field
intensities is qualitatively similar to the usual one discussed in
the Fano model. The spectra move towards lower energies with the
increasing pump-field intensities and cross at certain intensity
the energy of Fano zero, as documented in Figs.~12,~13. Whereas
the isolated atom $ b $ has a one-peak spectrum, the spectrum of
atom $ b $ influenced by atom $ a $ consists of two peaks that
form a spectral dublet at greater pump-field intensities clearly
visible in Fig.~13.

\subsection{Ionization spectra formed by the dipole-dipole interaction}

\begin{figure}[tbp]
\center{
\includegraphics[width=6.0cm]{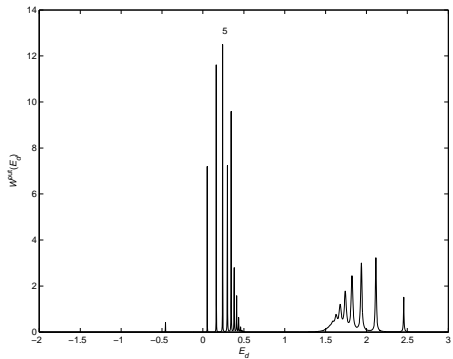}\hskip0mm
\caption{Photoelectron spectra $W^{\rm out}(E_d)$. The parameters
$E_a=E_b=E_L=1$, $q_a=100$, $\gamma_a=10^{-4}$, $q_b=\gamma_b=1$, $\Omega=0.1$,
$q_{\rm trans}=0$. Initially the laser mode is in a coherent state
with the mean photon number equal to 5.}
}
\end{figure}

\begin{figure}[tbp]
\center{
\includegraphics[width=6.0cm]{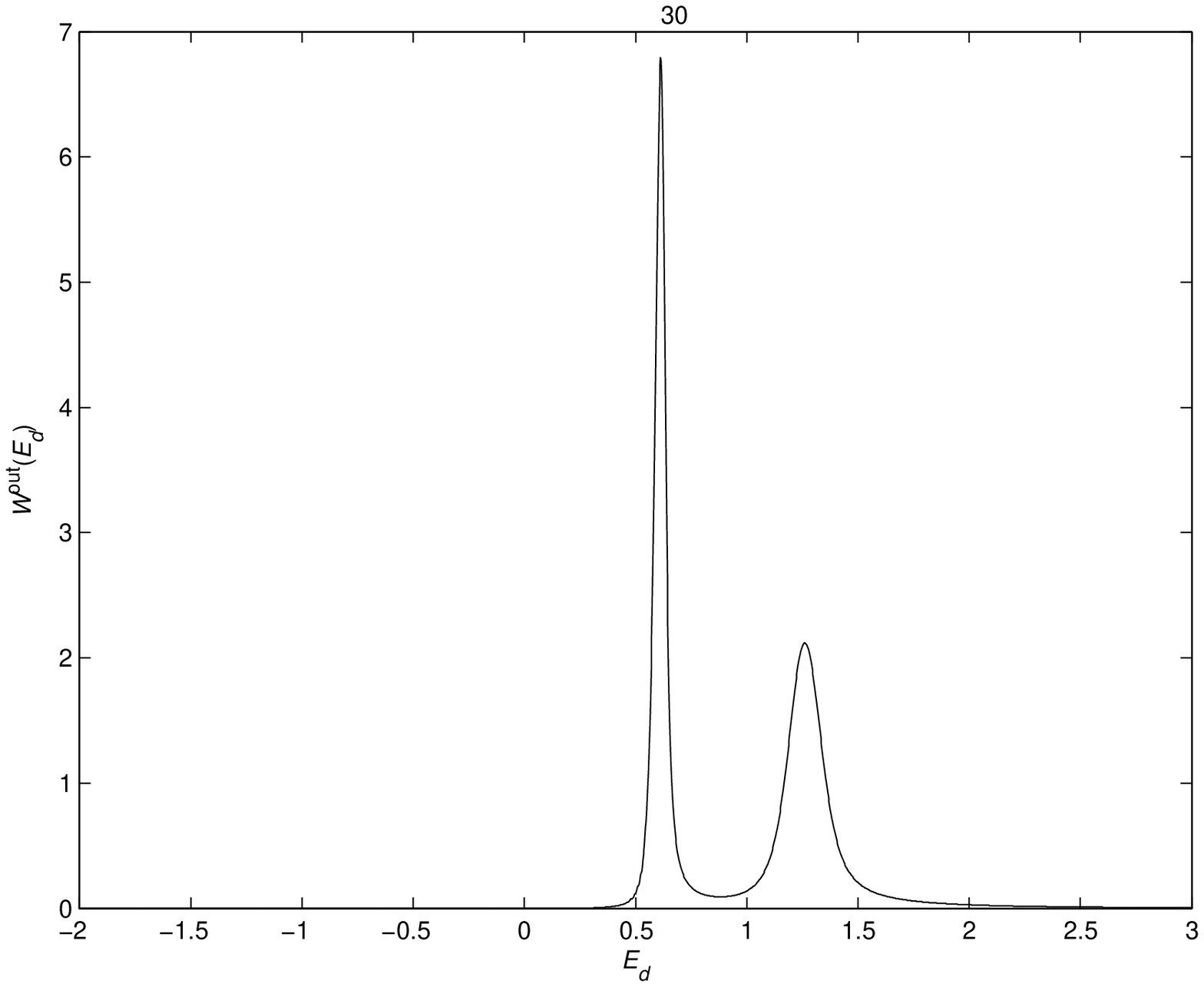}\hskip0mm
\caption{Same as in Fig.~14, but with the mean photon number equal to 30.}
}
\end{figure}

\begin{figure}[tbp]
\center{
\includegraphics[width=6.0cm]{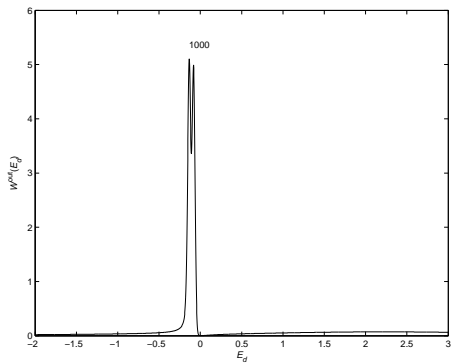}\hskip0mm
\caption{Same as in Fig.~14, but with the mean photon number equal to 1000.}
}
\end{figure}

The dipole-dipole interaction between atoms $ a $ and $ b $, in
general, splits the peaks in the ionization spectra of isolated
atom $ b $ into two parts (see Figs.~14,~15,~16). As a
consequence, there occur two major peaks in the ionization spectra
for greater pump-field intensities. These spectral peaks are
broken into many sub-peaks for low pump-field intensities as a
consequence of quantum character of the pump field (see Fig.~14).
Individual sub-peaks can be connected with the appropriate Fock
states, similarly as in the previous section. Two major peaks
approach each other with the increasing pump-field intensity and
form a spectral dublet at certain moment (see Fig.~15).

\begin{figure}[tbp]
\center{
\includegraphics[width=7.0cm]{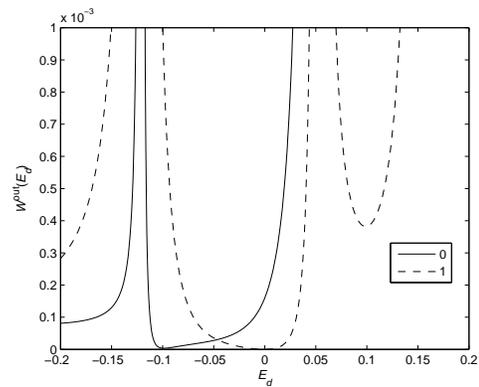}\hskip0mm
\caption{Photoelectron spectra $W^{\rm out}(E_d)$. Initially the
laser mode is in a coherent state with the mean photon number
equal to 3. The parameters $E_a=E_b=E_L=1$, $q_a=100$,
$\gamma_a=10^{-4}$, $q_b=\gamma_b=1$, $\Omega=0.1$, $q_{\rm
trans}=0\mbox{ (solid curve)}, 1\mbox{ (dashed curve)}$.} }
\end{figure}

If only the dipole-dipole interaction between the discrete level
of atom $ a $ and the continuum of states of atom $ b $ ($ J \ne 0
$) is considered, the Fano zero of isolated atom $ b $ is
inevitably lost. However, when also the dipole-dipole interaction
between the discrete levels of atoms $ a $ and $ b $ occurs ($
J_{ab} \ne 0 $), the Fano zero can be preserved under certain
conditions found in \cite{PLPeLe11a}. The two mentioned
dipole-dipole interactions compete in ionizing atom $ b $ in
certain sense. If the strengths of two interactions equal for the
energy of Fano zero formed at atom $ b $, the Fano zero is
preserved. The appropriate condition was derived in
\cite{PLPeLe11a} for the semi-classical model in the form
\begin{equation}   
 \frac{J_{ab}}{J} = \frac{\mu_b}{\mu}.
\label{44}
\end{equation}
Numerical computations have revealed that the condition in
Eq.~(\ref{44}) is valid also in the analyzed quantum model (see
Fig.~17). Here, we would like to note that the original Fano zero
of isolated atom $ b $ is usually replaced by a broad deep minimum
in the ionization spectra provided that the condition in
Eq.~(\ref{44}) is not fulfilled (see Fig.~17). Such behavior
originates in the weakness of dipole-dipole interactions compared
to the Coulomb and optical dipole interactions that form the Fano
zero of isolated atom $ b $.

\subsection{Entanglement of atoms $ a $ and $ b $}

We have assessed the entanglement by the 'computable measure of
entanglement', i.~e., the negativity \cite{VW02}. It is recommended
as such in the case of two parties (components) each possessing a
finite number of levels. We mark a difference, because in our
analysis one of the two parties has an infinite number of levels.
The straightforward approach was successful on the assumption of a
classical light field \cite{LPLePe12}, because the two components
are in a joint pure quantum state. To our knowledge, such an
approach cannot be based on simple formulas on inclusion of the
quantum nature of the field which leads to a mixed quantum state
describing the involved parties. Numerical calculation would be a
challenging task.

Recently, a selection of the frequencies has been realized in a
somewhat arbitrary, but systematic, way \cite{LPLePe12}. Two
states with these frequencies are just the levels needed for
producing a qubit. In such a way, we return to the well-known
two-qubit problem. For $E_d,E'_d$ $\in$ $[-2,3]$, $[-20,20],[-5,10],
[-1.5,1.5]$, we generate a 'density' plot of the negativity at $t=2$ that is
\begin{equation}
{\cal N}(t)
=\sum_{l=1}^4\frac{|\bar{\lambda}_l(t)|-\bar{\lambda}_l(t)}{2},
\label{45}
\end{equation}
where $\bar{\lambda}_l(t)$ are eigenvalues of the partially
transposed statistical matrix
{\footnotesize
\begin{equation}
\left(\begin{array}{cccc} \rho_{00|}^{\rm
out}(E_d,E_d,t)&\rho_{00|}^{\rm out}(E_d,E'_d,t)&
\rho_{10|}^{\rm out}(E_d,E_d,t)&\rho_{10|}^{\rm out}(E_d,E'_d,t)\\
\rho_{00|}^{\rm out}(E'_d,E_d,t)&\rho_{00|}^{\rm out}(E'_d,E'_d,t)&
\rho_{10|}^{\rm out}(E'_d,E_d,t)&\rho_{10|}^{\rm out}(E'_d,E'_d,t)\\
\rho_{01|}^{\rm out}(E_d,E_d,t)&\rho_{01|}^{\rm out}(E_d,E'_d,t)&
\rho_{11|}^{\rm out}(E_d,E_d,t)&\rho_{11|}^{\rm out}(E_d,E'_d,t)\\
\rho_{01|}^{\rm out}(E'_d,E_d,t)&\rho_{01|}^{\rm out}(E'_d,E'_d,t)&
\rho_{11|}^{\rm out}(E'_d,E_d,t)&\rho_{11|}^{\rm out}(E'_d,E'_d,t)
\end{array}
\right).
\label{46}
\end{equation} }
Here
\begin{eqnarray}
&& \mbox{} \left(\begin{array}{cc}
\rho_{jk|}^{\rm out}(E_d,E_d,t)&\rho_{jk|}^{\rm out}(E_d,E'_d,t)\\
\rho_{jk|}^{\rm out}(E'_d,E_d,t)&\rho_{jk|}^{\rm out}(E'_d,E'_d,t)\end{array}
\right)
\nonumber\\
&& \mbox{} = \frac{1}{\sum_{j=0}^1[\rho_{jj}^{\rm out}(E_d,E_d,t)
+\rho_{jj}^{\rm out}(E'_d,E'_d,t)]}
\nonumber\\
&& \mbox{} \times\left(\begin{array}{cc}
\rho_{jk}^{\rm out}(E_d,E_d,t)&\rho_{jk}^{\rm out}(E_d,E'_d,t)\\
\rho_{jk}^{\rm out}(E'_d,E_d,t)&\rho_{jk}^{\rm
out}(E'_d,E'_d,t)\end{array} \right),
\label{47}
\end{eqnarray}
with
\begin{eqnarray}
&& \mbox{}\rho_{jk}^{\rm out}(E_d,E_d,t)=\rho_{jk}^{\rm
out}(E_d,E'_d,t) (E'_d\rightarrow E_d),
\nonumber\\
&& \mbox{} \rho_{jk}^{\rm out}(E'_d,E_d,t)=\rho_{jk}^{\rm out}(E_d,E'_d,t)
(E_d\leftrightarrow E'_d),
\nonumber\\
&& \mbox{} \rho_{jk}^{\rm out}(E'_d,E'_d,t)=\rho_{jk}^{\rm out}
(E_d,E'_d,t)(E_d\rightarrow E'_d)
\mbox{, }
\nonumber\\
&& \mbox{} j,k=0,1.
\label{48}
\end{eqnarray}
Both the dipole-dipole interaction and the interaction mediated by
the quantized pump field create the entanglement between the bound
electron at atom $ a $ and the ionized electron at atom $ b $.
Suitable conditions for creating highly entangled states have been
revealed in \cite{LPLePe12} concerning a classical pump field. It
holds that the stronger the dipole-dipole interaction, the more
entangled state is reached. However, also a weaker dipole-dipole
interaction can provide highly entangled states provided that the
ionization process is sufficiently slow. This can be reached when
the strengths of the direct ionization path (connecting the ground
state of atom $ b $ with the continuum) and the indirect
ionization path (that ionizes an electron from the ground state of
atom $ b $ through the auto-ionizing discrete state of atom $ b $)
are balanced.

\begin{figure}[tbp]
\center{
\includegraphics[width=7.0cm]{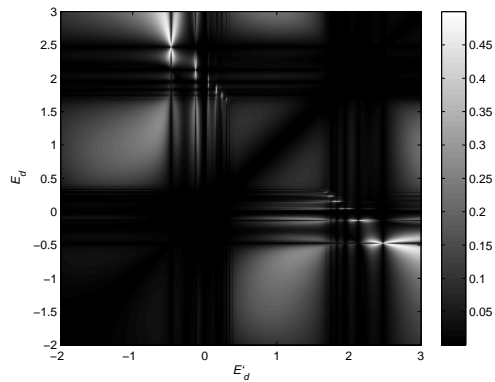}\hskip0mm
\caption{Density plot of the negativity ${\cal N}$ $(t=2)$ that measures
the entanglement between the neighbor atom $a$ and the atom $b$
with a continuum. The parameters $E_a=E_b=E_L=1$,
$q_a=100$, $q_b=\gamma_b=1$, $\Omega=0.1$, $q_{\rm trans}=0$.
Initially, the laser mode is in a coherent state with the
mean photon number equal to 1, $\gamma_a=0$.}
}
\end{figure}

\begin{figure}[tbp]
\center{
\includegraphics[width=7.0cm]{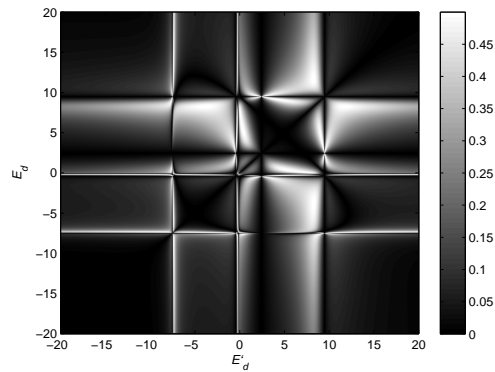}\hskip0mm
\caption{Same as in Fig.~18, but initially, the laser mode is in the Fock state
$|2\rangle_L$, $\gamma_a=10^{-4}$.}
}
\end{figure}

Similarly as in the semiclassical model analyzed in
\cite{LPLePe12}, the overall negativity can roughly be composed of
negativities of qubit-qubit systems obtained from the qubit of
atom $ a $ and all possible qubits found in the continuum of atom
$ b $. Such densities of negativity give us information about the
spectral distribution of entanglement. The density of negativity
for the ionization spectrum shown in Fig.~6 and appropriate for
the interaction mediated by the quantized field is plotted in
Fig.~18. We can see in Fig.~18 that the negativity is distributed
in the whole area of energies present in the ionization spectrum.
It is remarkable that the values of density of negativity are very
low for the degenerate energies of qubits inside the continuum of
atom $ b $ ($ E_d \approx E'_d $). This behavior can be explained
by the long-time energy conservation that does not allow to
entangle such qubits in the continuum with the qubit of atom $ a
$. The densities of negativity appropriate to the coherent and
Fock states completely differ, as demonstrated in Figs.~18,~19. We
note that a pump field in the Fock state with one photon cannot
create entanglement due to the energy conservation. However,
higher-number Fock states are already suitable for the
entanglement creation.

\begin{figure}[tbp]
\center{
\includegraphics[width=7.0cm]{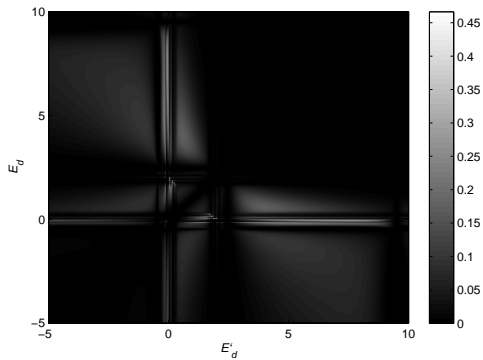}\hskip0mm
\caption{Same as Fig.~18, but $\gamma_a=10^{-4}$.}
}
\end{figure}

\begin{figure}[tbp]
\center{
\includegraphics[width=7.0cm]{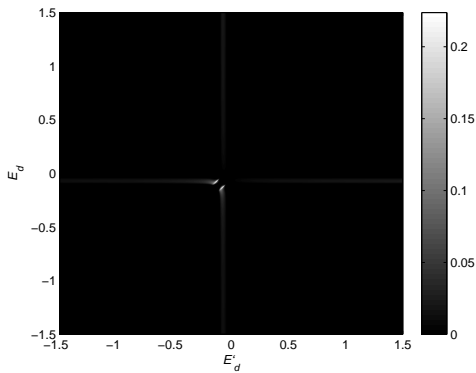}\hskip0mm
\caption{Same as Fig.~18, but the initial mean photon number is
equal to 1000, $\gamma_a=10^{-4}$.}
}
\end{figure}

The densities of negativities formed by the dipole-dipole
interaction behave similarly as those created by the interaction
mediated by the quantized field. It holds also here that
appreciable values of the density of negativity are found for
energies appreciably present in the ionization spectra. Also very
low values of the density of negativity occur around the
degenerate energies $ E_d \approx E'_d $ (see Figs.~20,~21). Thus,
the spectral concentration of negativity is observed as the
pump-field intensity increases (compare Figs.~20, 21). When the
spectrum forms a spectral dublet, the entanglement is encoded
between the two peaks of the dublet.

As follows from the above results, effects stemming from quantum
features of the pump optical field are clearly visible both in
ionization spectra and entanglement provided that the one-photon
'excitation' parameter $ \Omega $ is greater or comparable to 0.1
and the mean number of photons is smaller or comparable to 10.
Both coherent laser fields and highly-nonclassical Fock-state
fields are suitable for the observation of quantum signatures of
auto-ionization process. As for the Fock-state fields, they can be
generated, e.g., in heralded single-photon sources
\cite{Brida2012} or their generalizations \cite{PerinaJr2013} and
in QED cavities \cite{Varcoe2004}. Greater values of the
one-photon 'excitation' parameter $ \Omega $ represent
experimental challenge as the values reached in current ionization
experiments are much smaller. However, modern photonic band-gap
structures \cite{Joannopoulos2011,Sakoda2005} give a hope here.
They allow to dramatically increase electric-field amplitudes
inside due to constructive interference on one side. On the other
side, they form photonic bands with continuum of states which are
similar to those participating in ionization.

\section{Conclusions}

We have studied quantum correlations of two atoms. We have assumed
that one atomic system contains an auto-ionizing level whereas the
other atom does not comprise any auto-ionizing level. Both the
atoms interact with the same mode of the quantized field. We have
concentrated ourselves to the long-time behavior of the atomic
systems. The long-time behavior exhibits quantum correlations of
the two atoms even in the case where the atoms do not interact
directly. We have illustrated quantum correlations comparing the
one-peak spectrum appropriate for the neighbor atom without
optical excitation with the two-peak spectrum occurring for the
optically-excited neighbor atom. In the classical limit of strong
field the differences vanish. {\bf We have identified conditions
for the observation of quantum features in long-time electron
ionization spectra.} Also the Fano zero has been found in these
spectra for the quantized optical field considering the same
conditions as for the classical optical field.

\acknowledgements The authors acknowledge the financial support by
the project Operational Program Research and Development for
Innovations - European Social Fund (project
CZ.1.05/2.1.00/03.0058) of the Ministry of Education, Youth and
Sports of the Czech Republic and the project IGA No. PrF-014-05.

\end{document}